\documentclass[a4paper,twocolumn,english,aps,prx,reprint,superscriptaddress,showpacs,showkeys,longbibliography]{revtex4-1}
\usepackage[latin9]{inputenc}
\usepackage{babel}
\usepackage{mathtools}
\usepackage{amsmath}
\usepackage{amssymb}
\usepackage{graphicx}
\usepackage{float}
\usepackage[colorlinks=true, linkcolor=red, citecolor=blue, urlcolor=blue]{hyperref}
\usepackage[usenames,dvipsnames]{color}
\usepackage{bm}
\usepackage{bbold}
\usepackage{braket}
\usepackage{comment} 
\usepackage{dcolumn,extarrows}
\usepackage{soul}
\usepackage{footmisc}

\definecolor{mygreen}{rgb}{0,0.5,0}
\definecolor{myblue}{rgb}{0,0,0.75}
\definecolor{mymagenta}{cmyk}{0,1,0,0.12}

\newcommand{\minus}{
  \setbox0=\hbox{-}
  \vcenter{
    \hrule width\wd0 height \the\fontdimen8\textfont3
  }%
}

\def\inner(#1,#2,#3,#4,#5,#6){\ensuremath\left(\begin{array}{ccc} #1 & #2 & #3 \\ #4 & #5 & #6 \end{array}\right)}

\def\innerv(#1,#2,#3,#4,#5,#6){\ensuremath\left\{\begin{array}{ccc} #1 & #2 & #3 \\ #4 & #5 & #6 \end{array}\right\}}

\usepackage{babel}
\usepackage{braket}
\definecolor{mygreen}{rgb}{0,0.5,0}\definecolor{myblue}{rgb}{0,0,0.75}\definecolor{mymagenta}{cmyk}{0,1,0,0.12}

\makeatother

\begin{document}

\title{`Quantum Spin Lenses' in Atomic Arrays}

\author{A. W. Glaetzle}

\affiliation{Institute for Theoretical Physics, University of Innsbruck, A-6020
Innsbruck, Austria}

\affiliation{Institute for Quantum Optics and Quantum Information of the Austrian
Academy of Sciences, A-6020 Innsbruck, Austria}

\affiliation{Centre for Quantum Technologies, National University of Singapore,
3 Science Drive 2, Singapore 117543}

\affiliation{Clarendon Laboratory, University of Oxford, Parks Road, Oxford OX1
3PU, United Kingdom}

\author{K. Ender}

\affiliation{Institute for Theoretical Physics, University of Innsbruck, A-6020
Innsbruck, Austria}

\affiliation{Institute for Quantum Optics and Quantum Information of the Austrian
Academy of Sciences, A-6020 Innsbruck, Austria}

\author{D.~S.~Wild}

\affiliation{Department of Physics, Harvard University, Cambridge, Massachusetts
02138, USA}

\author{S.~Choi}

\affiliation{Department of Physics, Harvard University, Cambridge, Massachusetts
02138, USA}

\author{H.~Pichler}
\affiliation{ITAMP, Harvard-Smithsonian Center for Astrophysics, Cambridge, Massachusetts 02138, USA}
\affiliation{Department of Physics, Harvard University, Cambridge, Massachusetts
02138, USA}

\author{M. D. Lukin}

\affiliation{Department of Physics, Harvard University, Cambridge, Massachusetts
02138, USA}

\author{P. Zoller}

\affiliation{Institute for Theoretical Physics, University of Innsbruck, A-6020
Innsbruck, Austria}

\affiliation{Institute for Quantum Optics and Quantum Information of the Austrian
Academy of Sciences, A-6020 Innsbruck, Austria}
\date{\today}
\begin{abstract}
  We propose and discuss `quantum spin lenses', where quantum states of delocalized spin excitations in an atomic medium are `focused' in space in a coherent quantum process down to (essentially) single atoms. These can be employed to create controlled  interactions in a quantum light-matter interface, where photonic qubits stored in an atomic ensemble are mapped to a quantum register represented by single atoms. We propose Hamiltonians for quantum spin lenses as inhomogeneous spin models on lattices, which can be realized with Rydberg atoms in 1D, 2D and 3D, and with strings of trapped ions. We discuss both linear and non-linear quantum spin lenses: in a non-linear lens, repulsive spin-spin interactions lead to focusing dynamics conditional to the number of spin excitations. This allows the mapping of quantum superpositions of delocalized spin excitations to superpositions of spatial spin patterns, which can be addressed by light fields and manipulated. Finally, we propose multifocal quantum spin lenses as a way to generate and distribute entanglement between distant atoms in an atomic lattice array. 
\end{abstract}
\maketitle

\section{Introduction}

In quantum information processing \cite{ladd2010quantum} with atoms,
qubits are typically represented by internal atomic states, e.g.~as
long-lived spin excitations within the atomic ground state manifold~\cite{gardiner2015quantum2}.
Ideally, qubits are stored in single atoms, and for these qubits to
be identifiable and addressable, we typically require localization
of the atoms in well-defined spatial regions. Spatial control, and
localization of single atoms is a pre-requisite to implement single
and two-qubit operations, allowing addressing of individual qubits
with laser light, and providing entangling operations between adjacent
qubits by finite range interactions. Recent atomic physics experiments have demonstrated in a remarkable way the
basic ingredients of single atom manipulation and addressing for trapped
atoms and ions, and controlled interaction and entanglement between
atomic spin qubits with Rydberg atoms \cite{saffman2010quantum,Bloch2012},
trapped ions \cite{schindler2013quantum,debnath2016demonstration,jurcevic2014quasiparticle,richerme2014non},
cavity QED setups~\cite{reiserer2015cavity}, and quantum interfaces
\cite{kimble2008quantum,northup2014quantum,hucul2015modular}.

In contrast to localized qubits stored in single trapped atoms, atomic
ensembles provide us with qubits in the form of \emph{delocalized spin excitations}~\cite{Lukin2003,Hammerer2010}.
Delocalized spin qubits arise naturally in light-atomic ensemble interfaces
in both free space and cavity assisted setups. Here incident photons
representing a `flying qubit' are absorbed in an atomic ensemble with
enhanced interactions benefiting from a large atom number $N$, as
in an optically thick medium, and converted into a spin excitation,
which may be delocalized over the whole atomic cloud~\cite{Eisaman2005,Lamata2011,Peyronel2012,Heinze2013,Maxwell2013}.
In order to create controlled interactions between such delocalized qubits it is desirable to convert delocalized spin qubits into
localized qubits in the atomic array representing quantum memory.
Thus ideally we want operations --- \textit{a lens for spin
excitations} --- on the atomic array, which allow in a
coherent process `focusing' of qubits to a well-defined and localized
region, and ultimately to a single atom.

\begin{figure*}[tb]
  \includegraphics[width=1\textwidth]{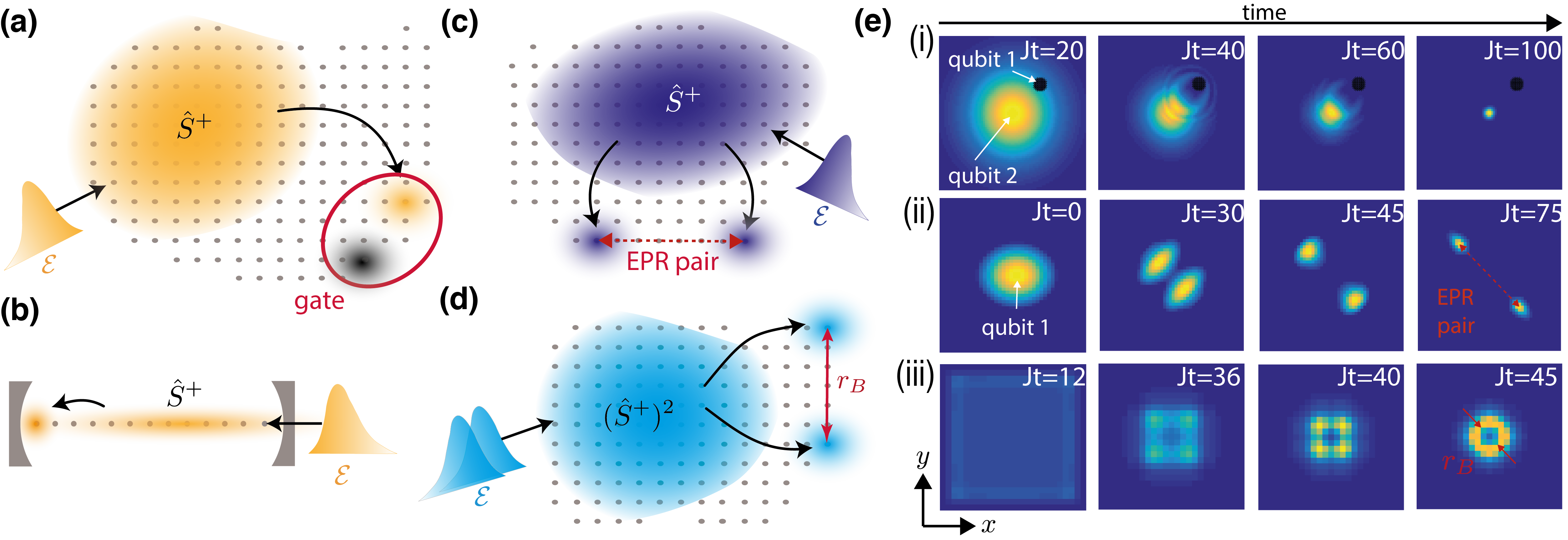} \caption{ (a -- d) Various scenarios of a quantum light-atom interface illustrating a \textit{quantum spin lens}. Incident photonic qubits are {initially stored} as delocalized spin excitations in an atomic array, and focused to single atoms (see Sec.~\ref{sec:linear}). (a) Basic process of write and focusing operations (see Secs.~\ref{sec:linear}A,B). Qubits stored in single atoms allow quantum gate operations to be implemented between adjacent atoms, e.g.~{via} Rydberg gates. (b)  1D setup with atomic ensemble stored inside a cavity (see Sec.~\ref{sec:linear}D). (c) Generation of EPR type states by constructing lenses with multiple focal points (see Sec.~\ref{sec:multifocal}).
(d) Focusing dynamics with a \textit{non-linear spin lens} with repulsive spin-spin interactions with range $r_B$, illustrated for two spin excitations (see Sec.~\ref{sec:nonlin}). 
(e) Density plots illustrating focusing dynamics of spin excitations in 2D arrays as a function of time $t$ according to spin-lens Hamiltonians \protect\eqref{eq:Hlattice} or \protect\eqref{eq:Hnonlin}. (i) Focusing of an initially delocalized `qubit $2$' in presence of a hole (representing, e.g.~a previously stored `qubit $1$') (see Sec.~\ref{sec:linear}). (ii) Two-focus lens illustrating generation of EPR pairs (see Sec.~\ref{sec:linear}C). Parameters: $50\times50$ lattice, initial width of the Gaussian wave packet $\sigma_{0}=10a$ lattice spacings. (iii) Focusing of two spin excitations with a non-linear quantum lens with repulsive spin-spin interactions of range $r_B$. The two excitations are focused to a ring, reminiscent of a quantum crystal. Parameters: $31 \times 31$ lattice. (see Sec.~\ref{sec:nonlin}).
}
\label{fig:setup} 
\end{figure*}

In this paper we propose and discuss {\em  linear and nonlinear `quantum spin lenses'} and their physical realization in quantum optical setups. We will first identify Hamiltonians to realize {\em linear spin lenses}, which map in a coherent process a delocalized to localized spin excitation, and {\em vice versa}. This has immediate application as a  quantum atom-light interface, where incident photonic qubits are sequentially stored in an atomic array, and focused to a quantum register of spatially localized spin qubits represented by single atoms [see Fig.~\ref{fig:setup}(a)]. Moreover, we can generalize the concept of the `quantum spin lens' to a \textit{multifocal lens}. In particular, this allows a single delocalized spin excitation to be mapped to a spatial EPR-like superposition state, thus providing a way to distribute or generate entanglement between (distant) atoms [c.f.~Fig.~\ref{fig:setup}(c)].  Finally, we will discuss the design of {\em non-linear spin lenses}, adding finite range (repulsive) spin-spin interactions to the spin-lens Hamiltonian. Thus focusing dynamics will be conditional on the number of initial spin excitations, and an initial quantum  superposition state of delocalized spins will be mapped to a superposition of spatial spin patterns [c.f.~Fig.~\ref{fig:setup}(d)]. Remarkably, this provides a tool to manipulating the individual terms (corresponding to a specific excitation number) in the superposition state by spatially addressing in the atomic medium. As noted above, the relevant spin models are naturally implemented in existing atomic and solid state quantum optical setups, and we will illustrate this below with the examples of neutral atoms with Rydberg-mediated spin-spin  interactions in 1D, 2D and 3D atomic lattices \cite{Maller2015,Labuhn2016,Zeiher2016,Jau2016} using laser-dressing techniques ~\cite{Henkel:2010ila,Pupillo:2010bta,Glaetzle2015,vanBijnen2015}, as well as with strings of trapped ions \cite{jurcevic2014quasiparticle,richerme2014non}.

\section{Linear Quantum Spin lenses: Focusing Dynamics of Single Spin Excitations}\label{sec:linear}

We are interested here in a scenario illustrated in Fig.~\ref{fig:setup}(a), where an incident wave packet~$\cal E$, representing a qubit $\alpha \ket{0}+\beta\ket{1}$ as a superposition of vacuum and a one-photon wave packet, is stored as a delocalized spin excitation in a medium of $N$ two-level atoms. These two-level systems can be physically represented by long-lived atomic hyperfine ground states two-level atoms $\ket{g},\ket{e}$, with all atoms initially prepared in the ground state, and we assume atoms trapped in an array.  Storage of a photonic qubit in the atomic medium is achieved, for example, in a Raman process ~\cite{Novikova2007,Gorshkov2007} , where the incident photon is absorbed and atoms, initially prepared in $\ket{g}$ transferred to $\ket{e}$. Writing to atomic quantum memory thus corresponds to a mapping of the photonic qubit to the atomic state $\alpha \ket{G}+\beta \hat S^{+} \ket{G}$. Here $\hat S^{+}=\sum_{n=1}^{N}\psi_{n}\hat \sigma_{+}^{(n)}$, a sum of Pauli raising operators for atoms $n$, creates a delocalized excitation distributed over the atoms according to an amplitude $\psi_{n}$, acting on the `vacuum state' $|G\rangle\equiv|g_{1},\ldots,g_{N}\rangle$ with all atoms in the ground state. This mapping of photons to spin excitations should be understood in the spirit of the Holstein-Primakoff approximation, where excitations in the atomic medium are essentially bosonic for small excitation fractions.

A quantum spin lens aims to achieve a mapping of the delocalized atomic spin excitation to (ideally) a  single atom, $\hat S^{+} \ket{G} \rightarrow \hat \sigma_{+}^{(n_f)}\ket{G}$ in a coherent quantum process, and preserving the superposition character of the qubit. 
Below we will first discuss spin-lens Hamiltonians that focus initially delocalized single spin excitations during the associated unitary time evolution.
We will call this focusing dynamics of single excitations \textit{linear} spin lenses, with \textit{nonlinear} spin lenses as focusing of multiple interacting spin excitations to be discussed in the following section. 

The focusing of single spin excitations discussed below can be generalized immediately to $k$ photonic qubits, provided we store and focus them \textit{sequentially}, i.e.~incident photonic qubits are absorbed and focused in the atomic medium one by one in spatially separated atoms $n_1,\ldots, n_\nu$ representing a quantum register~\footnote{ Sequential mapping of qubits requires transfer of qubits stored in the excited states $\ket{e}$ to another excited state $\ket{e^\prime}$ to \textit{hide} these qubits from the focusing dynamics of the following qubits. Note that these previously stored qubits appear as \textit{holes} (defects) in the focusing dynamics of the consecutive qubits, as discussed in Sec.~\ref{sec:defects}.}. Due to the spatial localization, these atomic qubits can now be individually addressed, and we can operate on them with single and two-qubit gate operations, implemented, for example, as Rydberg gates [see Fig.~\ref{fig:setup}(a) and panel~($i$) of Fig.~\ref{fig:setup}(e)].

\begin{figure*}[t]
\centering \includegraphics[width=1\textwidth]{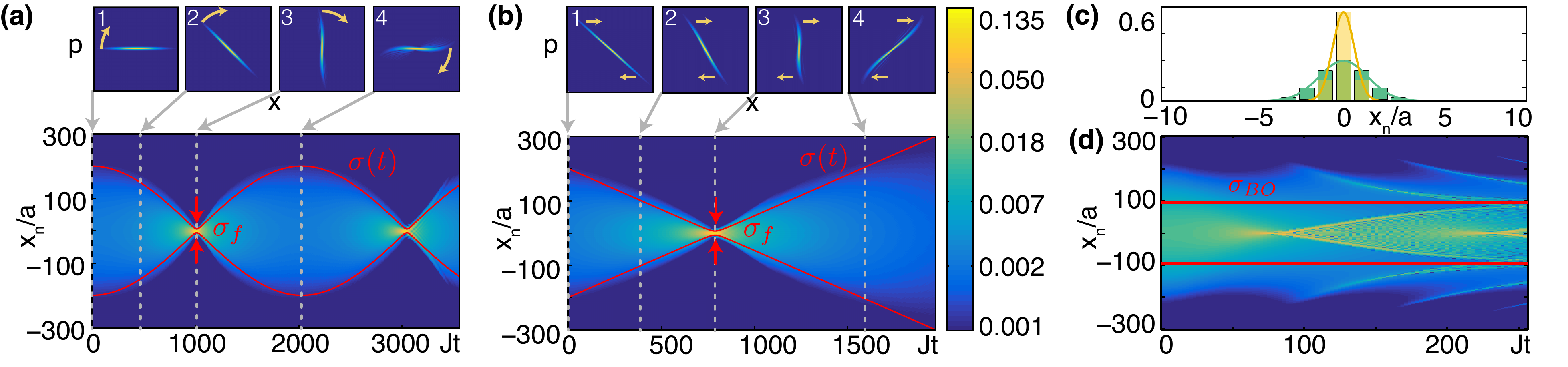}
\caption{{\small{}{}{Time evolution of an initially delocalized spin excitation on a lattice according to the spin-lens Hamiltonian \protect\eqref{eq:Hlattice} for the (a) }}\textit{\small{}{}thick}{\small{}{}
and (b) }\textit{\small{}{}thin}{\small{}{} lens for a 1D string
of $L=800$ spins as a function of time $t$. The red line, $\sigma(t)$,
indicates the continuum result. Insets 1-4 above illustrate the dynamics
of the Wigner function at four specific times.  (c)
Zoom to illustrate the final optimized spatial excitation probability
per atom illustrating that focusing on the scale of (essentially)
single atoms can be obtained. The green bars correspond to a focusing
dynamics of the thick lens with initial width $\sigma_{0}=100a$ and
final width $\sigma_{f}=2.7a$. The yellow bars correspond to a second
focusing stage with a thin lens starting with $\sigma_{0}=2.7a$ to
final width $\sigma_{f}=1.2a$. (d) Bloch oscillations in focusing
dynamics at the edge $\sigma_{\rm BO}$ of
the lens (see text).}}
\label{fig:dynamics} 
\end{figure*}

\subsection{Spin lens Hamiltonian with nearest-neighbor `flip-flop' interactions}

Focusing of a delocalized excitation in a spin chain is achieved with the Hamiltonian 
\begin{equation}
\hat H=-J\sum_{n}\left[\hat\sigma_{+}^{(n)}\hat\sigma_{-}^{(n+1)}+\textrm{H.c.}\right]+\sum_{n}V_{n}\hat\sigma_{z}^{(n)},\label{eq:Hlattice}
\end{equation}
where $\hat\sigma_{\mu}^{(n)}$ are Pauli spin $1/2$ operators at lattice
site~$n$. The first term describes hopping of the spin excitation
(kinetic energy), which for the moment we take as nearest neighbor
hopping, while the second term is a spatially dependent energy shift
$V_{n}=v_{0}(n-n_{f})^{2}$. While in Eq.~\eqref{eq:Hlattice} we write a 1D model, the present discussion can in a straightforward way be generalized to higher dimensions. 

The Hamiltonian \eqref{eq:Hlattice}  is motivated by analogy to
an optical lens with $V_{n}$ imprinting a  phase on the
$n$-th spin centered around lattice site $n_{f}$, reminiscent of
the refractive material of a lens~\cite{lohmann1993image,schleich2015quantum}. The analogy to an optical lens is best illustrated by visualizing the focusing dynamics of single spin excitations with a Wigner phase space distribution as a function of time [see Fig.~\ref{fig:setup}(e) and Fig.~\ref{fig:dynamics}(a,b), upper panel]. We write the wave function of the single spin excitation as \mbox{$|\psi(t)\rangle=\sum_n\psi_n(t)\hat\sigma_+^{(n)}|G\rangle$} with amplitude $\psi_{n}(t)$ initially delocalized as a wave packet of spatial width $\sigma_0$ over the lattice, and we define a Wigner function on the lattice as
 ~\cite{berry1977semi, bizarro1994weyl} ($\hbar=1$)
\begin{equation}
W_{\rm lat}(x_n,k)=\frac{a}{\pi}\int_{-\pi/2a}^{\pi/2a} dq\langle k-q|\psi\rangle \langle \psi|k+q\rangle e^{-2iqx_n} .
\nonumber
\end{equation}
Here  $x_n=an$ ($n\in\mathbb{Z}$) are discrete lattice positions with $a$ the lattice spacing, and momentum $k a \in(-\pi,\pi)$ is $2\pi$-periodic, and we denote by \mbox{$|k\rangle = ({{a}/{2\pi}})^{1/2}\sum_n e^{ikx_n}\hat\sigma_+^{(n)}|G\rangle$} spin waves with momentum $k$ on an infinite lattice. A momentum space representation of the time-dependent Schr\"odinger equation with Hamiltonian~\eqref{eq:Hlattice} shows that the dynamics is the one of a quantum pendulum. The first term in~\eqref{eq:Hlattice} gives rise to a Bloch band dispersion relation $\epsilon(k)=2J[1-\cos(ka)]$, and the quadratic potential term maps to a Laplacian in $k$, i.e.~a kinetic energy term.

Focusing dynamics is best illustrated in the continuum limit, i.e.~we assume that the spin dynamics is smooth on the scale of the lattice spacing, and the wave function in momentum space remains localized to a region at the bottom of the Bloch band. Thus the dispersion relation is well approximated by \mbox{$\epsilon(k)\approx J(ka)^{2}+\mathcal{O}(ka)^{4}$} for small momenta $ka\ll1$, and the Wigner function $W_{\rm lat}(x_n,k)$ maps to the standard Wigner function $W(x,p)$ for the continuous variables $x_n\rightarrow x\in\mathbb{R}$ and $k\rightarrow p\in\mathbb{R}$. The Hamiltonian~\eqref{eq:Hlattice} becomes
an effective harmonic oscillator~(HO), $H=p^{2}/(2m)+m\omega^{2}x^{2}/2$
with momentum $p$ and position $x$. Here we have defined
a frequency $\omega = 2\sqrt{v_{0}J}$, mass $m = 1/(2Ja^{2})$,
and we denote by $\ell=(\hbar/m\omega)^{1/2}$ the HO length. 
Under this Hamiltonian an initial Wigner function, $W(x,p;0)$, simply performs
a (classical) rigid rotation in phase space, $W(x,p;t)=W(\bar{x}(x,p,t),\bar{p}(x,p,t);0)$,
where position, $\bar{x}(x,p,t)=m\omega x\cos\omega t-p\sin\omega t$,
and momentum, $\bar{p}(x,p,t)=p\cos\omega t+m\omega x\sin\omega t$,
describe elliptical trajectories in phase space as a function of time. 
Thus, an initial wave packet with width $\sigma_0$ in position space is transformed after a quarter of period, $t_{f}=\pi/(2\omega)$, to a spatially localized state with width $\sigma_f=\ell^2/\sigma_0\ll \ell$, as familiar from squeezed states \cite{gardiner2015quantum2}  \footnote{In contrast to focusing with \textit{quench dynamics} in a HO, as described above, one could also localize the wave function in an adiabatic ramp of the harmonic oscillator, i.e.~by sweeping  $\omega \rightarrow \omega_f$. An initial Gaussian wave packet, which is matched to represent the HO ground state with $\omega$ with width $\ell$ would then be mapped to the final ground state with width $\ell_f\equiv \sqrt{\hbar /m \omega_f}$, with time required $\gg 1/\omega$. 
{\it Quench} dynamics discussed in this paper results in fast focusing $t_{f}\sim 1/\omega$, and  does not require good knowledge of the initial wave function or control over the applied trapping potential to match the initial wave packet to the HO ground state. In contrast, an adiabatic scheme can be expected to be more robust against imperfect parameters in Hamiltonian ~\eqref{eq:Hlattice}\label{fn:adi}}.

In this continuum approximation the single particle Schr\"odinger equation from~\eqref{eq:Hlattice} is formally equivalent to the paraxial Helmholtz equation~\cite{schleich2015quantum}. The role of time in the Schr\"odinger equation is replaced by the axial dimension in the paraxial Helmholtz equation, and the potential translates to a spatially dependent refractive index. This allows to interpret most of the focussing dynamics in the language of classical optics. So far we considered focusing of a delocalized excitation in a potential $V_n$, which is `on' during the whole dynamics. In analogy to optics this corresponds to light propagating in a graded index multimode fiber. In the following we will refer to this dynamics as a `thick lens'.  This is to distinguish from a second scenario, discussed below, where focussing is achieved by a `thin lens'.

In Fig.~\ref{fig:dynamics}(a) we illustrate  focusing
dynamics of a `thick lens' for spin excitations with the lattice model
\eqref{eq:Hlattice} in a parameter regime where the continuum approximation
is well justified (see below for details). For an initial Gaussian
wave packet with spatial width~$\sigma_{0}$, corresponding to a (in the continuum limit)
cigar shaped Wigner function 
\mbox{%
$W(x,p,0)=(1/\pi)\exp[{-(x/\sigma_{0})^{2}-(\sigma_{0}p)^{2}}]$%
}, the spatial width %
\mbox{
$\sigma(t)^{2}=\sigma_{0}^{2}[\cos^{2}\omega t+\left({\ell}/{\sigma_{0}}\right)^{4}\sin^{2}\omega t]$
} starts oscillating as a function of time {[}see red line in panel~(a){]} and has periodic minima at every quarter of a period $\omega t_{f}=\pi/2$
where $\sigma_{f}\equiv\sigma(t_{f})=\ell^{2}/\sigma_{0}$. The final
width in real space (after a quarter of a period) corresponds to the
Fourier transform of the initial wavefunction, i.e.~$\psi(x,t_{{\rm foc}})=\mathcal{F}_{x'}\{\psi(x',0)\}(x/\ell^{2})$,
as for an optical lens. The focusing in real space is illustrated in Fig.~\ref{fig:dynamics}, where panels~1-4 show the corresponding phase space dynamics of the Wigner function.

Instead of the `always-on' Hamiltonian \eqref{eq:Hlattice} of the `thick lens', focusing can also be obtained in a pulsed scheme,
where the quadratic potential term is switched on for a short time
only. This imprints a
position dependent momentum kick $\Delta ka=-2\phi_{0}(n-n_{f})$ (with $\phi_0>0$) onto
the initial wavefunction via the quadratic phase shift %
\mbox{%
  $\hat U_{\phi_{0}}=\exp[-i\phi_{0}\sum_{n}(n-n_{f})^{2}\hat \sigma_{z}^{(n)}]$%
},  followed by a free evolution of the spin system via \eqref{eq:Hlattice}
with $v_{0}=0$, as illustrated in Fig.~\ref{fig:dynamics}(b). This is in analogy with a `thin lens' in classical
optics, where a thin refractive material imprints a position dependent
phase onto the incoming plane wave. The Wigner function in phase space first acquires a momentum
kick $p\rightarrow p-2\phi_{0}x/a^{2}$, followed by free evolution
corresponding to a shear motion of the Wigner function, i.e.~$W(x,p,t)=W(x-{pt}/{m},p,0)$, as
illustrated in panels~1-4 of Fig.~\ref{fig:dynamics}(b). In contrast
to the `thick lens' there is a single focal time $Jt_{f}=2(\sigma_{0}/a)^{4}\phi_{0}/[4\phi_{0}^{2}(\sigma_{0}/a)^{4}+1]$ ($\approx1/(2\phi_0)$ for $\sigma_0 \gg a$)
where a Gaussian wavefunction has its minimum width $\sigma_{f}=\sigma_{0}/\sqrt{4\phi_{0}^{2}(\sigma_{0}/a)^{4}+1}$.

Figure~\ref{fig:dynamics}(c) shows the spatial
excitation probability for lattice sites around the focus. The green
bars correspond to the dynamics illustrated in panel~(a). An excitation
initially delocalized over $\sigma_{0}/a=100$ lattice sites gets
localized on $\sigma_{f}/a=2.7$ lattice sites using a `thick' lens including corrections up to sixth order (described in \ref{sec:linear}{\color{red}B}). One can improve the focusing by using multiple pulses or even combining the two different schemes. For example, we can further focus the spin excitation from $\sigma_{0}/a=2.7$ lattice
sites to $\sigma_{f}/a=1.2$ lattice sites, shown as
the yellow bars in panel~(c), by adjusting the lens strength to the new initial condition.

\begin{figure}[t]
\centering \includegraphics[width=.85\columnwidth]{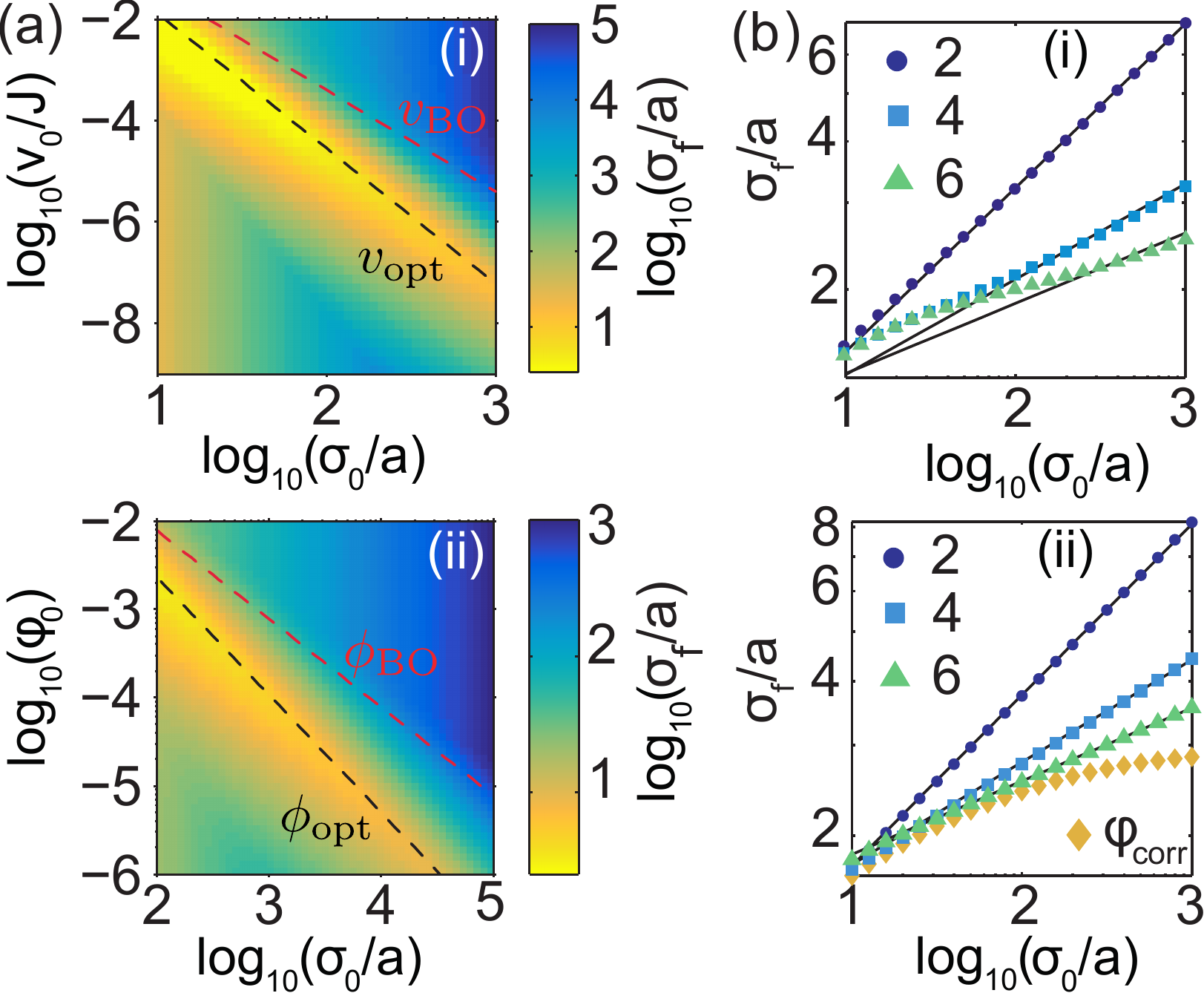}
\caption{{\small (a) Final width $\sigma_{f}$ as a function of initial width
$\sigma_{0}$ and (i) potential strength $v_{0}$ or (ii) phase $\phi_{0}$
for the (i) `thick' and (ii) `thin' lens setup. Dephasing due
to the non-quadratic dispersion relation starts to dominate at $v_{0}=v_{\rm opt}(\sigma_{0})$
and $\phi_{0}=\phi_{\rm opt}(\sigma_{0})$ (black dashed line) while Bloch
oscillations start at $v_{0}=v_{\mathrm{{BO}}}(\sigma_{0})$ and $\phi_{0}=\phi_{\mathrm{{BO}}}(\sigma_{0})$ (red dashed line) for `thick' and `thin' lenses, respectively. (b) Minimum final width $\sigma_f$ for the optimized lens strength as a function of initial width for (i) `thick' and (ii) `thin' lens setup. Dark blue dots, light blue squares and green triangles correspond to the numerically obtained final width for a quadratic, quartic, and sixth order potential. The black lines are a guide to the eye, showing the respective scalings of Eq.~\eqref{eq:scalingopt}.}}
\label{fig:corr} 
\end{figure}

\subsection{Lattice Corrections \textendash{} Dephasing and Bloch Oscillations}\label{sec:corr}

Corrections to the continuum limit become important when the delocalized excitation is focused to a spatial region on the order of the lattice spacing, and the Wigner function extends close to the border of the first Brillouin zone. This happens  for  `sufficiently strong potential' in Eq.~\eqref{eq:Hlattice}, which
leads to aberration and Bloch oscillations due to deviation from a
quadratic dispersion relation. In the following we discuss the limitations this imposes on the achievable final width and show how the effects of the non-quadratic dispersion relation can be compensated using non-parabolic lens potentials.

The main results are summarized in Fig.~\ref{fig:corr}(a)
 where the numerically obtained final width $\sigma_{f}$ is plotted as a function
of the initial width $\sigma_{0}$ and lens strength $v_{0}$ or $\phi_{0}$
for the (i) `thick' and (ii) `thin' lens on a lattice, respectively. Simulations have been
performed on a 1D chain with $L=10^6$ ($L=10^5$) spins according to
Eq.~\eqref{eq:Hlattice} for the thin (thick) lens setup. In contrast
to the continuum picture, where a stronger lens leads to a tighter
spatial focus and a faster focusing time, the numerical results show
that there exist optimal `lens potentials' (see below) scaling as
\begin{equation}
v_{\rm opt}\sim\left(\frac{a}{\sigma_{0}}\right)^{8/3}\quad\text{and}\quad
\phi_{\rm opt}\sim\left(\frac{a}{\sigma_{0}}\right)^{4/3},\\
\label{eq:optimum}
\end{equation}
for the `thick' and `thin' lens setups, illustrated as black dashed lines in Figs.~\ref{fig:corr}(a). At this optimal lens strength the final achievable width scales as
\begin{equation}
\sigma_{f}=a\kappa\left(\frac{\sigma_{0}}{a}\right)^{1/3}.
\label{eq:scaling}
\end{equation}
for both the `thick' and `thin' lens with $\kappa$ obtained numerically. Focusing works well below this optimal lens strength, in excellent agreement with Figs.~\ref{fig:dynamics}(a,b). The scaling of  Eq.~\eqref{eq:scaling}  (black line) is in perfect agreement with the numerically evaluated final width (blue dots) shown in Fig.~\ref{fig:corr}(b) for (i) the `thick' and (ii) the `thin' lens with $\kappa=0.68$ and $\kappa=0.80$, respectively.

In the following we discuss the two main effects of the lattice: abberation, giving rise to the optimal lens potentials of Eq.~\eqref{eq:optimum}, and Bloch oscillations at the edge of the lens shown in Fig.~\ref{fig:dynamics}(d). We show that the corresponding aberration can be addressed using potentials and pulse shapes that include quartic and higher order terms. 

{\it Abberation:}
Deviations from the continuum model can be understood as arising from the non-linear
group velocity $v_{g}(k)=2Ja\sin(ka)$ on the lattice, which implies that wave
packets with large momenta propagate more slowly in a lattice than in the continuum
limit, where $v_{g}^{(c)}(k)=2Ja^{2}k$. The difference between
the two velocities is only negligible provided the path difference during
the focusing time is small compared to the final size of the wave packet, i.e.
$[v_{g}^{(c)}(k)-v_{g}(k)]t_{f}\ll\sigma_{f}$. Expanding the sine up to third order and evaluating the equation at the maximum momentum $k\sim1/\sigma_f$ we obtain Eq.~\eqref{eq:optimum} (see appendix~\ref{app:correction}). The difference in the group velocities further explains the `$s$'-shaped distortion of the Wigner function observed panels~1-4 of Fig.~\ref{fig:dynamics}(a,b) as the non-linear group velocity induces a non-rigid rotation in phase space. 

{\it Bloch oscillations:}
At an even larger potential strength, the wings of the wave packet with an extension larger than
\begin{equation}
\sigma_{\rm BO}=2a\sqrt{\frac{J}{v_{0}}}
\label{eq:sbo}
\end{equation}
will undergo Bloch oscillations (see appendix~\ref{app:correction}), as
illustrated in Fig.~\ref{fig:dynamics}(d). This limits the lens strength to values well-below $v_{\mathrm{BO}}=4 J(a/\sigma_{0})^{2}$, indicated as  red dashed line in Figs.~\ref{fig:corr}(a).  At this critical lens potential the (local) potential gradient $V'_n= 2 v_0 (n-n_f)$ 
gives rise to Bloch oscillations with an amplitude $\Delta n =2J/V'_n$ and frequency $\omega_{\rm BO}=V_n'/2$~\cite{dahan1996bloch}. If this frequency becomes of the order of the focusing time focusing becomes ineffective.
A similar argument for the thin lens
yields the maximum pulse strength $\phi_{\rm BO}\sim a/\sigma_{0}$.
In contrast to Bloch oscillations in the thick lens, for the thin lens focusing is limited by phase wraps
as the momentum kicks imparted by the pulse extend beyond the first
Brillouin zone. 

{\it Correction of aberration:}
Quartic deviations from the dispersion relation limit the final width of
the spin wave to \mbox{$\sigma_{f}\sim (\sigma_{0}/a)^{1/3}$}. 
Similar to aspherical lenses, the effect of dephasing due to the non-quadratic
dispersion relation, can be compensated using more general potentials (`thick' lens) and imprinted phase
profiles (`thin' lens), of the form 
\begin{equation}
V_n=\sum_{q=1}^{Q/2} v_{2q} \;  (n-n_f)^{2q},
\label{eq:VnQ}
\end{equation}
including additionally quartic ($Q=4$), sixth ($Q=6$) or even higher order terms. Such higher order terms will accelerate the wings of
the wave packet stronger compensating for their smaller group velocities on the lattice. Using a similar argument as in appendix~\ref{app:correction}, one can show that for an appropriate choice of $v_{2q}$ this leads to an improved scaling
\begin{equation}
\sigma_{f}\sim \left(\frac{\sigma_{0}}{a}\right)^{1/(Q+1)}.
\label{eq:scalingopt}
\end{equation}
Fig.~\ref{fig:corr}(b) plots the final width as a function of the initial width for ($i$) the `thin' and ($ii$) the `thick' lens setup using a lens strength up to $Q=2,4$ and $6$. The numerically obtained final width agrees well with the scaling of Eq.~\eqref{eq:scalingopt}.

For the `thin' lens an optimized form of Eq.~\eqref{eq:VnQ} can be derived analytically using a semi-classical model with continuous spatial variable $an\rightarrow x$ and a Bloch band dispersion $\epsilon(k)$ giving rise to a non-linear group velocity $v_g(k)$. Given the imprinted phase profile $\phi(x)$, the initial wave packet receives a position dependent momentum kick $\phi'(x)/a$. In order to focus all parts of the wave packet to the focal point $x_f=0$ at the same time $t_f$, we require $x=v_g(k(\phi'))t_f$ which yields
\begin{equation}
\phi(x)=-(x/a)\arcsin[\phi_0 (x/a)]-\sqrt{\phi_0 ^2-(x/a)^2},
\label{eq:phicorr}
\end{equation}
with $t_f=1/(2J\phi_0)$. Panel~(ii) of Fig.~\ref{fig:corr}(b) shows the final width obtained using~\eqref{eq:phicorr} with $x=na$ on a lattice (yellow diamonds), which shows a clear improvement over the parabolic phase profile. Note that $\phi(x)$ is only real-valued up to $|x/a| = \phi_0$  due to the maximum group velocity $v_g(\pi/2) = 2J$ on a lattice, since only parts of the wave packets with distance $x<2Jt_f$ can be focussed within the focusing time.

\begin{figure}[bt]
\centering 
\includegraphics[width=1\columnwidth]{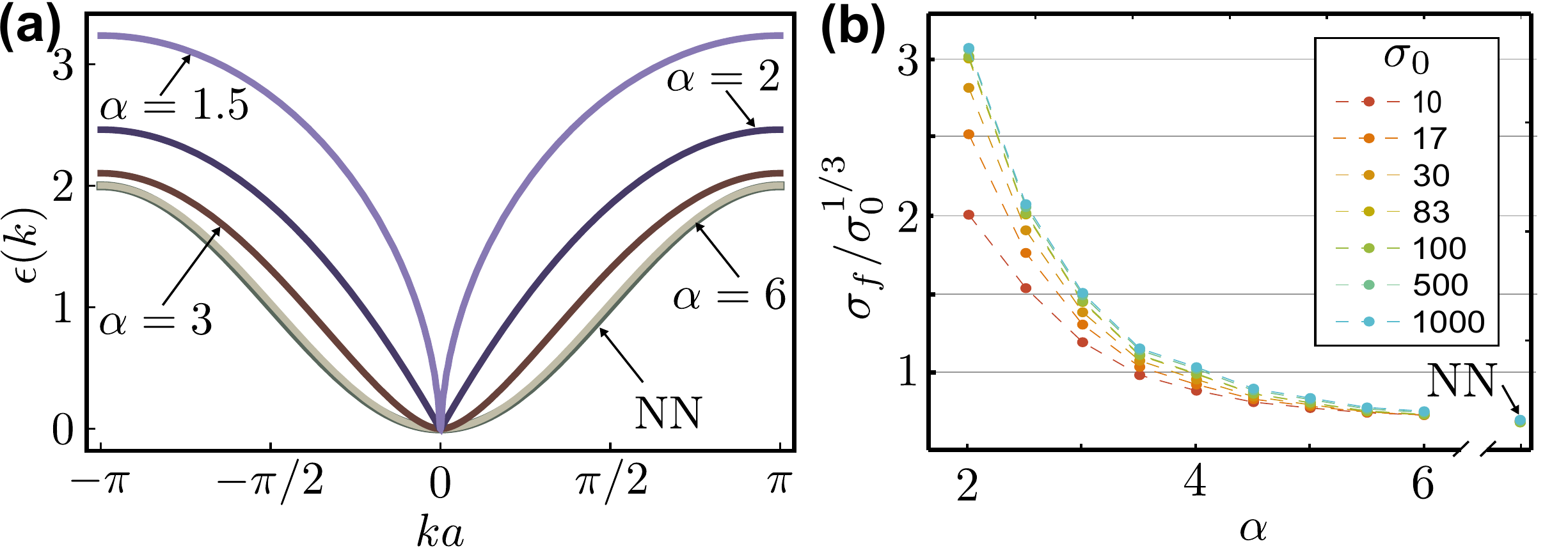} \caption{{\small (a) Dispersion relation $\epsilon_\alpha(k)$ for long-range `flip-flop' interactions (see text) for different exponents $\alpha$ and nearest-neighbor~(NN) interactions. (b)~Rescaled final width $\sigma_f/\sigma_0^{1/3}$ for different exponents $\alpha$ and for different initial conditions $\sigma_0$. The data points on the very right correspond to nearest-neighbor interactions.}}
\label{fig:nn_ato_comp} 
\end{figure}

\subsection{Multifocal Lenses and Generation of EPR states}\label{sec:multifocal}

Instead of the single focus lens, as in \eqref{eq:Hlattice}, we can employ double well, or multi-well potentials. 
The corresponding potentials can be generated as spin dependent optical potentials, and an array of spin lenses can be realized with large spacing optical lattices. A multifocal lens operating on a single initial delocalized spin excitation  will generate a superposition state of excitations at the focal points. For two foci, for example, we can generate an EPR type state
\begin{equation}\label{eq:EPR}
\hat S_{}^+|G\rangle\longrightarrow\left(\hat \sigma_+^{(n_1)}+\hat \sigma_+^{(n_2)}\right)|G\rangle. 
\end{equation}
Thus we generate a superposition (EPR state) between spins at lattice site~$n_1$ and~$n_2$, as schematically illustrated in panel~($ii$) of Fig.~\ref{fig:setup}(c). 

The time evolution of the corresponding multi-focal lens is visualized in panel~($ii$) of Fig.~\ref{fig:setup}(e). In the upper half plane   ($y_n>0$) we used the 2D potential \mbox{$V({x_n,y_m})=v_0[(x_n-x_f)^2+(y_m-y_f)^2]$} with focal points \mbox{$(x_f,y_f)=( 0 , \sqrt{2}\times10 )a$} while in the lower half plane ($y_n<0$) we used $V({x_n,y_m})$ with  \mbox{$(x_f,y_f)=( 0 , -\sqrt{2}\times10 )a$}. Note that in  Fig.~\ref{fig:setup}(e) we rotated the potential by 45 degrees. The potential strength is optimized to \mbox{$v_0=4.5\times 10^{-3}J$} for an initially symmetric  Gaussian wave function with radial spatial width \mbox{$\sigma_0=10 a$} on a 2D lattice with \mbox{$L_x=L_y=50$} lattice sites.

\subsection{Long-range  `flip-flop' interactions}

Implementations of Hamiltonian~\eqref{eq:Hlattice} with Rydberg atom in optical lattices or strings of trapped ions motivate a model 
\begin{equation}
\hat H=-\sum_{n,m}J_{n}\left[\hat\sigma_{+}^{(m)}\hat \sigma_{-}^{(m+n)}+\textrm{H.c.}\right]+\sum_{n}V_{n}\hat \sigma_{z}^{(n)},\label{eq:HlatticeLR}
\end{equation}
with \textit{long-range} `flip-flop' interactions $J_{n}=J_{0}/n^{\alpha}$. In particular, dipolar and van der Waals interactions between Rydberg (dressed) atoms allow to realize $\alpha=3$ and $\alpha=6$~\cite{saffman2010quantum}, respectively, while $0<\alpha<3$ can be realized with strings of ions~\cite{schindler2013quantum,debnath2016demonstration,jurcevic2014quasiparticle,richerme2014non}. The first term of~\eqref{eq:HlatticeLR} gives rise to a dispersion relation $\epsilon_\alpha(k)=2\sum_{n}[1-\cos(nka)]/n^\alpha$, as shown in Fig.~\ref{fig:nn_ato_comp}(a). While for $\alpha=6$ the dispersion relation $\epsilon_6''(0)=\pi^4/90\approx 1.08$ closely resembles the one of the nearest-neighbor `flip-flop' interactions of~\eqref{eq:Hlattice}, for $\alpha<3$ the dispersion relation exhibits a kink, e.g.  $\epsilon_2(k)=({\pi}/{2})|ka|-{(ka)^2}/{4}$, resulting in a linear group velocity at small momenta with a discontinuity at $k=0$. This leads to strong aberration and inefficient focusing. We note that this can be corrected using an adiabatic lens schemes~\footnotemark[2]. 

Fig.~\ref{fig:nn_ato_comp}(b) shows $\kappa\equiv\sigma_f/(a^2\sigma_0)^{1/3}$ [see Eq.~\eqref{eq:scaling}] for  different realizations of $\alpha$ for the `thick' lens setups.  While for large values of $\alpha$ the scaling of Eq.~\eqref{eq:scaling} agrees with the numerically obtained final width, for smaller values the linear dispersion relation leads to strong deviations. The smallest final width (smallest $\kappa$) is obtained for large values of $\alpha$ and ultimately with nearest-neigbor interactions, however, $\alpha=6$ almost perfectly resembles nearest-neighbor interactions.

\section{A Non-linear quantum spin lens: Focusing and spatial sorting of multi-photon
states}

\label{sec:nonlin} While the lenses discussed so far are \emph{linear
lenses} operating on single spin excitations, we can also design \emph{non-linear
lenses}, where the focusing dynamics depend on the number of spin excitations
in the medium via spin-spin interactions. Returning to the light-matter
interface discussed at the beginning of Sec.~\ref{sec:linear}, we
now generalize to an incident {\it multiphoton} superposition state \mbox{$|\mathcal{E}\rangle=\sum_{\nu=0}^{\infty}c_{\nu}|\nu\rangle$}.
For a write process to atomic quantum memory using a Raman scheme involving
a pair of atomic ground state levels (as described in Ref.~\cite{Fleischhauer2002}), this multiphoton state will be mapped to a superposition of (dilute) delocalized spin excitations, \mbox{$|\mathcal{E}\rangle \rightarrow\sum_{\nu}c_{\nu}(\hat{S}^{+})^{\nu}|G\rangle/\sqrt{\nu!}$} (representing hardcore bosons). Repulsive spin-spin interactions, which become relevant during the focusing dynamics when the excitation density increases, will map this superposition state to a
\textit{superposition of spatial spin patterns} in an atomic quantum
memory.  We note that this provides a means of manipulating the individual
terms in the superposition state by spatially addressing the atomic spins
with a laser. These transformed superposition states of spins can
then be mapped back to photons in a defocusing and read operation,
providing effective nonlinearities and manipulation of quantum states
on the single photon level.

\begin{figure}[tb]
\centering \includegraphics[width=1\columnwidth]{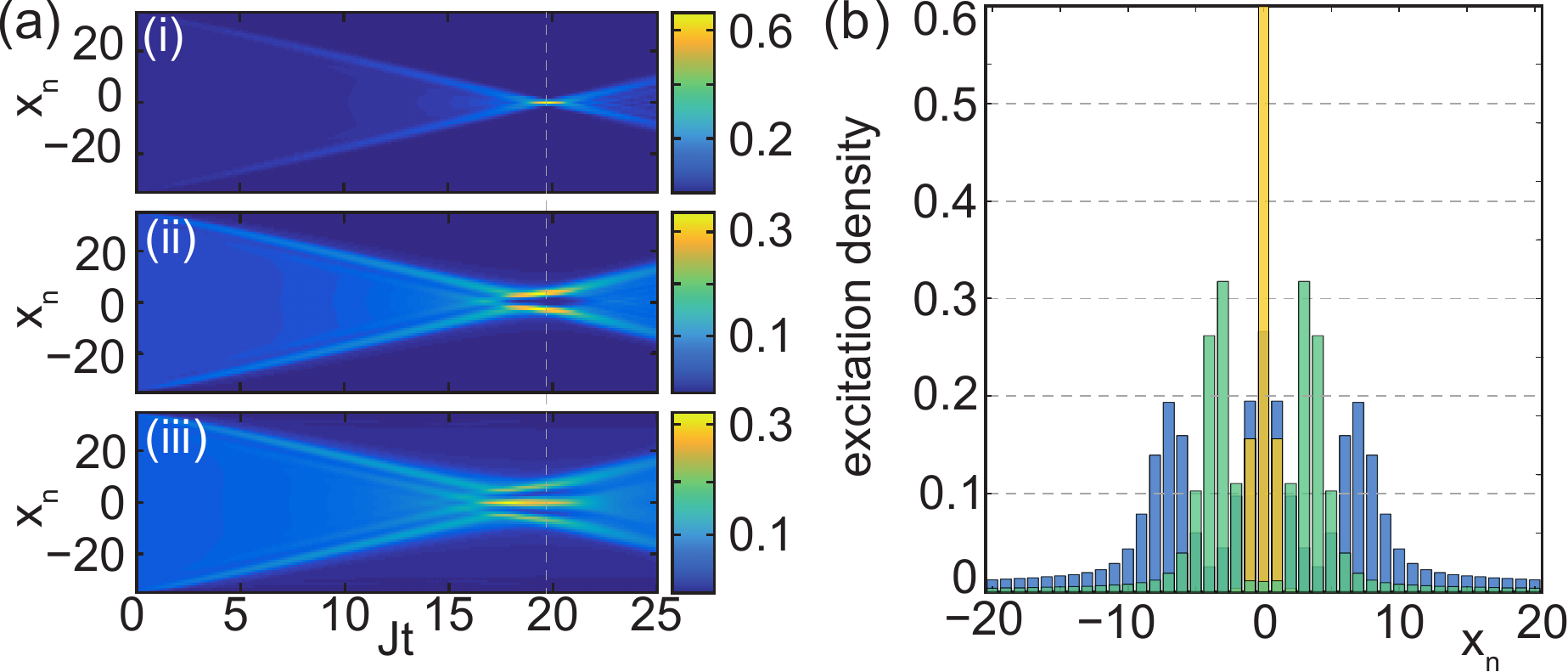}
\caption{{\small{}Time evolution of the excitation probability $p_{n}^{(\nu)}(t)$
for (i) $\nu=1$, (ii) $\nu=2$ and (iii) $\nu=3$ initially delocalized
spin excitation according to the non-linear spin-lens Hamiltonian
(\ref{eq:Hnonlin}) with interaction range $r_{B}=4.1a$ and $J_z=5\times 10^{3}J$ (b) Excitation probability $p_{n}^{(\nu)}(t_{f})$ at the focal time $t_{f}$ for $\nu=1$ (yellow), $\nu=2$ (green) and $\nu=3$ (blue) excitations, demonstrating spatial spin pattern formation depending on the number of excitations $\nu$.}}
\label{fig:nonlin}
\end{figure}

Non-linear quantum lenses can be implemented by generalizing the Hamiltonian
(\ref{eq:Hlattice}) to
\begin{multline}
\hat H=-J\sum_{n}\left[\hat{\sigma}_{+}^{(n)}\hat{\sigma}_{-}^{(n+1)}+\textrm{H.c.}\right]+\sum_{n}V_{n}\hat{\sigma}_{z}^{(n)}\\
+\sum_{n}\sum_{m}J_{z}^{(m)}\hat{\sigma}_{z}^{(n)}\hat{\sigma}_{z}^{(n+m)},\label{eq:Hnonlin}
\end{multline}
with the last term a long-range $J_{z}^{(m)}=J_z/m^6$ spin-spin interaction and blockade distance $r_B=a (J_z/J)^{1/6}$. We emphasize that the spin-spin interactions in~\eqref{eq:Hnonlin} arise naturally in Rydberg (dressed) gases and in trapped ion spin models.
Time evolution according to the above Hamiltonian will propagate the initial
quantum state to a  strongly  correlated many-body quantum state,
\begin{eqnarray}
 \sum_{\nu=0}^{\infty}c_{\nu} \frac{1}{\sqrt{\nu!}} \left(\hat{S}^{+}\right)^{\nu}|G\rangle\longrightarrow\sum_{\nu=0}^{\infty}c_{\nu}|\psi_{\nu}\rangle\label{eq:nonlin}
\end{eqnarray}
with
\begin{eqnarray}
|\psi_{\nu}\rangle=\sum_{n_{1},...,n_{\nu}}\psi_{n_{1},...,n_{\nu}}^{(\nu)}\hat{\sigma}_{+}^{(n_{1})}\ldots\hat{\sigma}_{+}^{(n_{\nu})}|G\rangle,\label{eq:psik}
\end{eqnarray}
and $\psi_{n_{1},...,n_{\nu}}^{(\nu)}$ the spatial wave functions for
$\nu$ spin excitations.

Figure~\ref{fig:nonlin} illustrates these focusing dynamics of interacting
spins according to~(\ref{eq:Hnonlin}) for an initial superposition
state consisting of exactly one, two or three
delocalized spin excitations as a function of time. We plot the excitation
probability as a function of position in the lattice, $p_{n}^{(\nu)}\equiv{\rm tr}\{\hat{\sigma}_{+}^{(n)}\hat{\sigma}_{-}^{(n)}|\psi_{\nu}\rangle\langle\psi_{\nu}|\}$
at lattice site $n$ for $\nu=1,2,3$, which clearly exhibits the spatial
mapping and resolution of spin patterns associated with $|\psi_{\nu}\rangle$
of Eq.~(\ref{eq:psik}). This allows to perform gate operations on spatially localized atoms, e.g. atoms $n=\pm 4$ or $n=\pm 7$, in order to manipulate the  $\nu=2$ or $\nu=3$ contribution of the superposition state individually. We note that the small excitation fraction between the peaks, e.g. population of atoms with $n=-1,0,1$ for $\nu=2$ (green bars), can be traced back to states in the initial wave function where two excitations were closer than $r_{B}$. This fraction of states becomes smaller by decreasing the initial excitation density, i.e. increasing the atom number.

The above can be immediately generalized to higher dimensions. In particular,
Fig.~\ref{fig:setup}(e), bottom panel~(iii), illustrates focusing
of two spin excitations ($\nu=2$) in 2D. In this case the repulsive spin-spin interactions give rise to a superposition of states with two excitations separated by a characteristic distance $r_B$ around the single-excitation focus forming a ring, reminiscent of a quantum crystal.

\section{Implementation with Rydberg atoms in 2D and 3D arrays}

The quantum spin lenses proposed in the previous sections can be implemented with atoms stored in optical trap arrays, including large spacing optical lattices and and optical tweezers~\cite{Endres2016,Barredo2016,maller2015rydberg} in 1D, 2D and 3D, or alternatively with trapped ions in 1D~\cite{schindler2013quantum,debnath2016demonstration,jurcevic2014quasiparticle,richerme2014non}. 

\begin{figure}[tb]
\includegraphics[width=1\columnwidth]{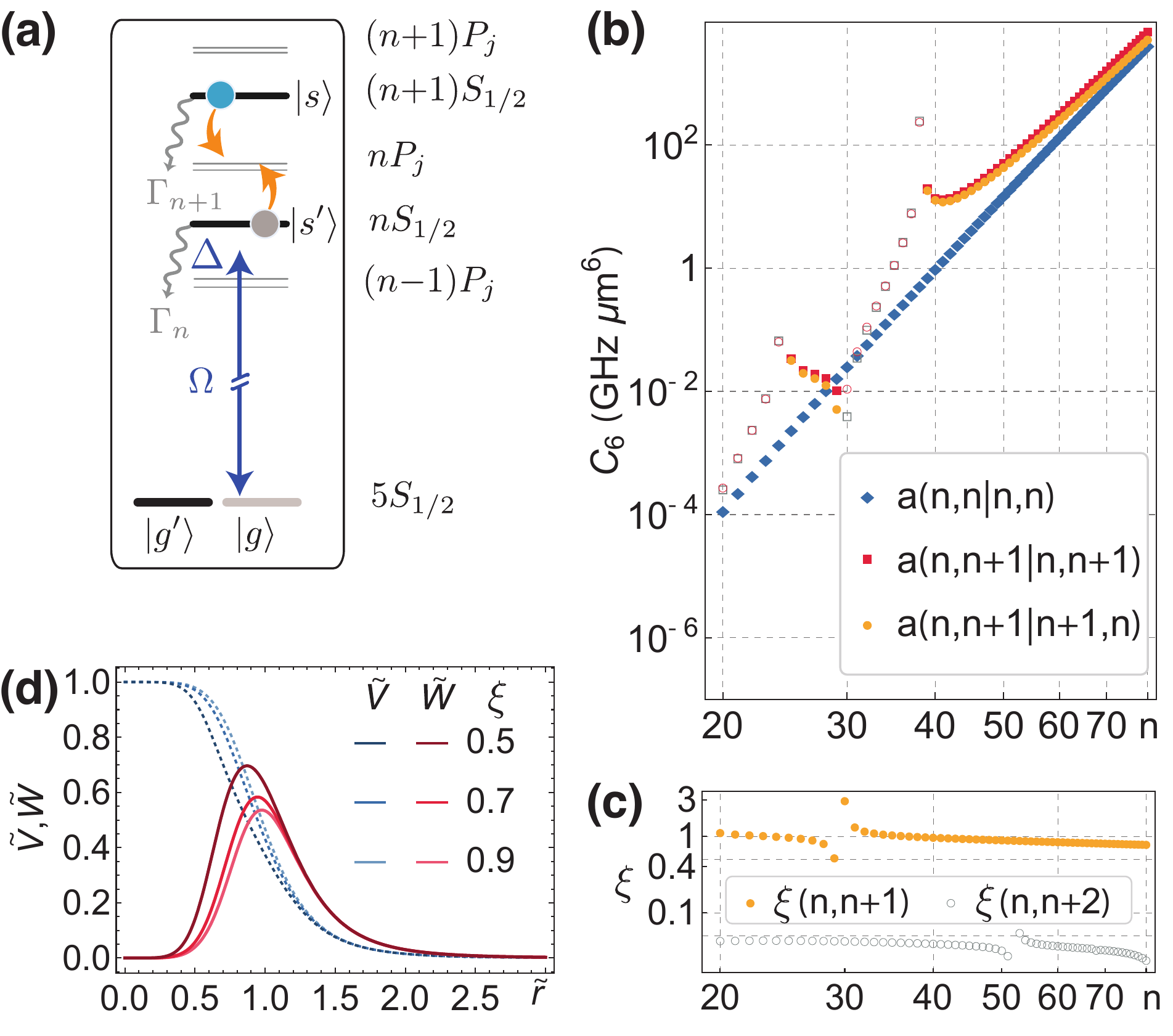}
\caption{{\small{}{}{
(a)  Atomic level scheme (see text).
(b) Van der Waals coefficients between Rydberg $nS_{1/2}$
and $n'S_{1/2}$ states as a function of principal quantum number
$n$. Negative values are plotted as outlined markers, positive values
as filled. 
(c) Exchange interaction strength $\xi(n,n+\delta n)\equiv a(n+\delta n|n,n+\delta n)/a(n,n+\delta n|n,n+\delta n)$ as a function of principal quantum number $n$. 
(d) Effective interaction potentials of Eq.~\eqref{eq:VWtilde} for $\xi=0.5$, 0.7 and
0.9, e.g., corresponding to }}\mbox{{$n=29,90$ and~27.}}}
\label{fig:Rydberg} 
\end{figure}

Below we describe first a realization of a linear spin-lens Hamiltonians of the type \eqref{eq:Hlattice} in 1D, but in particular also in 2D and 3D with alkali Rydberg atoms, where the spin degree of freedom of Sec.~\ref{sec:linear} is represented by a pair of levels involving a long-lived atomic hyperfine ground state, and a highly-excited Rydberg state. As an example, we consider $^{87}$Rb
atoms and choose $|g\rangle=|{5S_{1/2},F=2,m_{F}=2}\rangle$ as the
spin down and $|s\rangle=|nS_{1/2},m_{j}=1/2\rangle$ as the spin up
state [see Fig.~\ref{fig:Rydberg}(a)]. Note that in this section we denote by $n$ the principal quantum number.

Long-range spin exchange interactions %
\mbox{%
$J(r_{ij})=J/r_{ij}^{6}$%
} between spins $i$ and $j$ in 3D can be achieved by weakly dressing
the atomic ground state $|g\rangle$ by admixing with a laser a second Rydberg state %
\mbox{%
$|r\rangle=|n'S_{1/2},m_{j}=1/2\rangle$%
}  with (effective) Rabi frequency
$\Omega$ and detuning $\Delta\gg\Omega$. This particular choice of Rydberg
states leads to spin couplings $J(r_{ij})$, which are {\em  isotropic} in
3D, i.e.~a purely radial dependence as a function of the distance $r_{ij}=|\mathbf{r}_{i}-\mathbf{r}_{j}|$
for a large range of principal quantum numbers~\cite{bijnen2013,Hannaford2017}.
We note that, e.g.~dipolar exchange interactions can be engineered
by dressing the ground state with Rydberg $|n'P_{j},m_{j}\rangle$ states
resulting in anisotropic flip-flop interactions of the form %
\mbox{%
$J_{ij}\sim J/r_{ij}^{3}$%
}~\cite{Barredo2015}.

To obtain the desired flip-flop term in Eq.~\eqref{eq:Hlattice}
we first consider two atoms and derive an effective Hamiltonian for
the dynamics between the dressed ground state and the Rydberg state.
We start with a microscopic Hamiltonian, %
\mbox{%
$\hat H_{{\rm mic}}=\hat H_{A}^{(1)}+ \hat H_{A}^{(2)}+\hat H_{{\rm int}}$%
}, where the first two terms account for the two driven atoms with
\mbox{%
$\hat H_{A}^{(i)}=-\Delta|r\rangle_{i}\langle r|+(\Omega/2)|g\rangle_{i}\langle r|+{\rm h.c.}$%
} written in a rotating frame
. A small magnetic field and a circularly polarized laser
beam allows  dressing of the ground state with a specific Zeeman sublevel
of the Rydberg state.

The key element of the implementation is the van der Waals interaction,
$\hat H_{{\rm vdW}}$, between the $nS_{1/2}$ and $n'S_{1/2}$ Rydberg
states. Choosing two $s$-states ensures that the resulting vdW interactions
are isotropic in 3D over a large range of principal quantum numbers $n$ (see appendix~\ref{app:vdw}).
The exchange interaction between the degenerate states %
\mbox{%
$|nS_{1/2},1/2\rangle\otimes|n'S_{1/2},1/2\rangle$%
} and %
\mbox{%
$|n'S_{1/2},1/2\rangle\otimes|nS_{1/2},1/2\rangle$%
} dominantly arises via virtual population of $|(n-1)P_{j},m_{j}\rangle\otimes|(n'+1)P_{j'},m_{j}'\rangle$
Rydberg states [see Fig.~\ref{fig:Rydberg}(a)] and strongly depends on
$n$ and $n'$. As a particular example to demonstrate the tunability
of the resulting spin interactions we discuss the case $n'=n-1$ for
which the exchange process is maximized [see Fig~\ref{fig:Rydberg}(c)]. The interaction has the structure 
\begin{equation}
\hat H_{\text{vdW}}=\frac{1}{r^{6}}\left(\begin{array}{ccc}
c_{11} & 0 & 0\\
0 & c_{12} & w_{12}\\
0 & w_{12} & c_{12}
\end{array}\right),
\end{equation}
written in the basis of states $|r,r\rangle$, $|r,s\rangle$ and
$|s,r\rangle$ where we neglect the $\ket{s,s}$ interactions, since
we start initially with only one excitation and considering a linear lens. The generalized vdW coefficients $c_{ij}=a(n_{i},n_{j}|n_{i},n_{j})$
(diagonal) and $w_{ij}=a(n_{i},n_{j}|n_{j},n_{i})$ (exchange) are
shown in Figure~\ref{fig:Rydberg}(a) and derived in appendix~\ref{app:vdw}. The linear behavior (on the log-scale) of $c_{11}$ shows
the typical $n^{11}$ scaling of vdW interactions. The resonances
in $c_{12}$ and $w_{12}$ for $n=25$ and $n=40$ stem from vanishingly
small energy differences between to the states $\ket{nP_{3/2},nP_{1/2}}$
and $\ket{nP_{3/2},nP_{3/2}}$, respectively. Close to one of the
F\"orster resonances diagonal and off-diagonal interactions become
approximately equal, $c_{12}\approx w_{12}$, which are both dominated
by a single channel.

Adiabatic elimination of $\ket{r}$ in the limit $\Omega\ll\Delta$
leads to an effective long-range spin model between the dressed ground
state $\ket{g}$ and the Rydberg state $\ket{s}$ of the form 
\begin{equation}
\hat H_{{\rm eff}}=\sum_{ij}\left[V_{sg}^{(ij)}\hat \sigma_{gg}^{(i)}\hat\sigma_{ss}^{(j)}\!+\!\frac12 W_{sg}^{(ij)}\hat\sigma_{gs}^{(i)}\hat\sigma_{sg}^{(j)}\right]\label{eq:RydSpin}
\end{equation}
with Pauli operators $\hat \sigma_{ij}=|i\rangle\langle j|$ and effective
laser admixed interactions $V_{sg}^{(ij)}=\Omega^{2}/(4\Delta)\tilde{V}(r_{ij})$
and $W_{sg}^{(ij)}=\Omega^{2}/(2\Delta)\tilde{W}(r_{ij})$ given by
\begin{equation}
\tilde{V}=\frac{\tilde{r}^{12}+\tilde{r}^{6}}{(\tilde{r}^{6}+1)^{2}-\xi^{2}}\quad\text{and}\quad\tilde{W}=\frac{\xi\tilde{r}^{6}}{(\tilde{r}^{6}+1)^{2}-\xi^{2}}.\label{eq:VWtilde}
\end{equation}
Here, $\tilde{r}=(|\Delta|/c_{12})^{1/6}r$, ($\Delta<0$) is a dimensionless
distance and $\xi=w_{12}/c_{12}$ is the relative exchange strength
(see Fig.~\ref{fig:Rydberg}). Note that we have dropped the AC Stark shift, which affects all $s$ states equally. The potentials are shown in Fig.~\ref{fig:Rydberg}(a)
as a function of interatomic distance. As a particular example we consider dressing to the $n=60$ Rydberg state with (two-photon) Rabi frequency $\Omega/2\pi=10\text{MHz}$ and detuning $\Delta/2\pi=-20\text{MHz}$
with a lattice constant $a$ adjusted to the maximum of $\tilde{W}$
which results in $J/2\pi=0.36$~MHz resulting in a typical focusing
times for an initial width $\sigma_0=100a$ around $t_f=5\mu\text{s}$ which increases linearly with the initial width
This compares well with the lifetime of the state $\ket{s}=|60S_{1/2},m_{j}=1/2\rangle$
with $\tau_{60S}=252\mu\text{s}$~\cite{Beterov} resulting in $t_{f}/\tau\approx0.02$.

Instead of the spin models with atomic ground and Rydberg states representing a spin $1/2$  system, one can also employ dressing schemes, where spin is represented by a pair of long-lived and trapped atomic ground states, and spin hopping and interaction terms  are obtained by admixing with a laser Rydberg interactions ~\cite{Glaetzle2015,vanBijnen2015}. Such schemes may be convenient for the non-linear lenses described in Sec.~\ref{sec:nonlin}.

\section{Effects of disorder on focusing dynamics}
\label{sec:defects}

\begin{figure}[tb]
  \includegraphics[width=1\columnwidth]{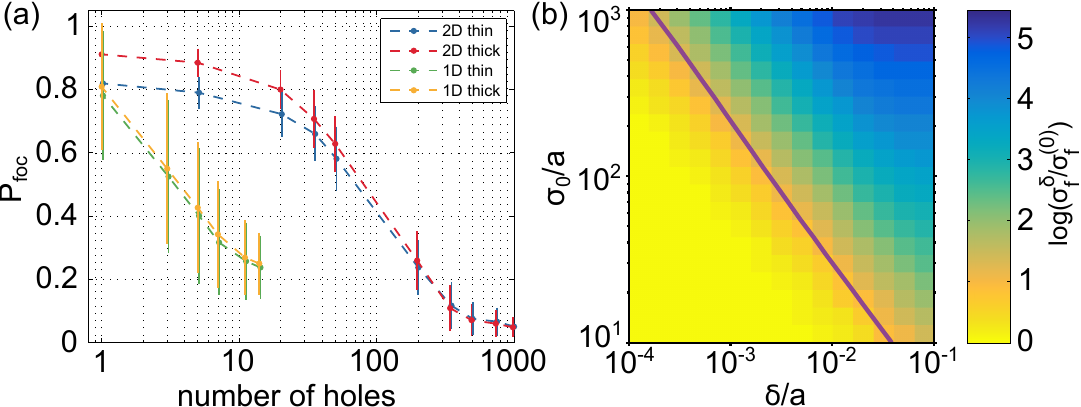} \caption{{\small{}{}{(a) Integrated probability $P_{\text{foc}}$ to find the focused spin wave inside a circle of radius 3 around the set focal point as a function of the number of holes in the lattice. (b) Final width $\sigma_f^\delta$ of a spin wave in a disordered system compared to the final width $\sigma_f^{(0)}$ without disorder as a function of the initial width $\sigma_{0}$ and the disorder strength $\delta$. The results were obtained for a thick lens including corrections up to sixth order (described in \ref{sec:linear}{\color{red}B}), where the potential strength and the focusing time were chosen to minimize $\sigma_f^{(0)}$ for each value of $\sigma_0$. The line represents $\delta \sim 1/ t_\text{foc}$, indicating the breakdown of focusing due to disorder (details in text).}}}
\label{fig:disorder} 
\end{figure}

In this section we analyze the robustness of the focusing dynamics against
two types of static disorder in the spin lattice: ($i$) holes in the lattice and ($ii$) static fluctuations of the atomic position resulting in a random distribution of long-range spin couplings $V^{(ij)}_{sg}$ and $W^{(ij)}_{sg}$ in Eq.~\eqref{eq:RydSpin}.

\textit{Non-unity filling:} 
Missing atoms in the lattice may arise due to imperfect loading or when a previously focused spin wave is stored in a different hyperfine ground state. In addition to being excluded from the hopping matrix $W_{sg}$, each hole is surrounded by an effective potential due to the modification of $V_{sg}$ in Eq.~\eqref{eq:RydSpin}. We numerically investigated the effect of randomly distributed holes in both one and two dimensions. Fig.~\ref{fig:disorder}(a) shows the spin excitation probability within a radius $3a$ around the focus of the lens, $P_{\text{foc}}$, for a 1D and 2D spin lattice of $N=70$ and $N=70\times70$ sites, respectively. Each data point is obtained by averaging over 1000 random hole realizations for the 1D example (400 realizations for 2D), starting with a Gaussian wavefunction with initial width $\sigma_{0}=14a$. The width of the statistical distribution of the final probability is indicated by the error bars.

For the 1D case, a single hole already has a significant detrimental effect on the final wave packet, which we attribute to the fact that $W_{sg}\sim1/r^{6}$ closely resembles nearest-neighbor hopping. By contrast, in 2D the focusing-scheme is almost unaffected by a small number of holes ($\lesssim 10$), as the spin wave can `flow' around the holes. This is further illustrated in panel~(i) of Fig.~\ref{fig:setup}(e) as a sequence of snapshots showing the 2D lattice of spins as a function of time. We expect the focusing scheme to be even more robust in three dimensions as there are more paths to avoid the holes. 

\textit{Static disorder in atomic positions:} As a second form of disorder we analyze the effect of fluctuations of the long-range spin couplings $V_{sg}$ and $W_{sg}$. Such models have previously been discussed in the context of Rydberg atoms trapped in optical tweezers \cite{Marcuzzi2017}. We assume that the position of the $n$-th atom is given by $\textbf{r}_{n}=\textbf{r}_{n}^{(0)}+\textbf{d}_{n}$, with $\textbf{r}_{n}^{(0)}$ the position on a regular lattice and the displacement $\textbf{d}_{n}\sim\mathcal{N}(0,\delta^{2})$ drawn from a normal distribution with zero mean and standard deviation~$\delta$. This results in a change of the interatomic separation $\textbf{r}_{n}-\textbf{r}_{m}=\textbf{r}_{nm}^{(0)}+\textbf{d}_{nm}$, in turn modifying the diagonal and hopping potentials of Eq.~\eqref{eq:RydSpin} to $W_{sg}(r_{ij})\approx W_{sg}(r_{ij}^{(0)})+ d_{ij}W'_{sg}(r_{ij}^{(0)})$ and $V_{sg}(r_{ij})\approx V_{sg}(r_{ij}^{(0)})+d_{ij}V'_{sg}(r_{ij}^{(0)})$. We note that the first order term in the expansion of $W_{sg}$ may vanish for certain separations, e.g. nearest neighbors, if the maximum of the interaction potential $\tilde W$ is commensurate with the lattice. However, the first order term will be present for all other separations, as well as in the expansion of $V_{sg}$ which exhibits no maximum or minimum as a function of distance (see Fig.~\ref{fig:Rydberg}). 

Disorder tends to localize the eigenstates of the system and thus prevents focusing when the localization length is smaller than the initial width of the wave packet~\cite{anderson1958}. However, the focusing fidelity may be significantly reduced even if the localization length is large. To quantify the role of disorder we estimate the energy broadening of plane wave states due to position disorder. In a regular lattice, plane wave states with momentum $k$ are energy eigenstates following the dispersion relation $\epsilon(k)$. In the presence of weak disorder, states with similar momenta are coupled together such that the energy of a plane wave acquires an uncertainty on the order of $\Delta \epsilon(k) = \sqrt{\bra{k} (\hat H - \hat H_0)^2 \ket{k}}$, where $\hat H$ and $\hat H_0$ denote the Hamiltonian of the disordered and the clean system, respectively. Since our scheme sensitively relies on the interference between different momentum states, we expect that focusing ceases to be effective when the focusing time exceeds $t_\text{foc} \gtrsim 1/\Delta \epsilon$. Given that $\Delta \epsilon \sim \delta$ to lowest order in $\delta$, this suggests that there exists a critical disorder strength $\delta_c \sim 1/t_\mathrm{foc}$ above which the disorder strongly affects the focusing dynamics. Indeed, this simple argument correctly predicts the breakdown of focusing as demonstrated in Fig.~\ref{fig:disorder}(b). We note that we numerically verified that the argument applies equally well to the `thin lens'.

\section{Conclusion and Outlook}

In this work we have shown that lenses for spin qubits can be designed
for atomic lattice gases, allowing focusing of delocalized spin excitations
in quench dynamics to essentially single atoms. In addition, we have provided an implementation
of a spin lens based on Rydberg-dressed spin-spin interactions. The
present work defines a novel light-matter interface,
where incoming photons are stored in delocalized atomic excitations
in an atomic medium, with spin focusing providing the link and mapping
to storage of qubits in single atoms. We note that existing experimental
setups with Rydberg atoms \cite{Maller2015,Labuhn2016,Zeiher2016,Jau2016} enabling the physical realization
of 1D and 2D $XY$-spin models can provide first proof-of-principle
experiments: here a single delocalized spin excitation as initial condition
could be generated using the Rydberg-blockade mechanism in an atomic
lattice~\cite{Schauss2012,Endres2016,Barredo2015}, and with focusing
dynamics implemented as described in the present work. This scenario
could also be demonstrated with the spin models realized with trapped ions \cite{jurcevic2014quasiparticle,richerme2014non}.
Finally, we expect that optimal coherent control techniques both for
spatial and temporal model parameters should allow for significant
improvement of `spin lenses'~\cite{caneva2011chopped}.

\textit{Acknowledgments:} We acknowledge discussions with M.~Heyl
and UQUAM partners C.~Gross and I.~Bloch. Work at Innsbruck is supported
by the Austrian Science Fund SFB FoQuS (FWF Project No. F4016-N23),
the European Research Council (ERC) Synergy Grant UQUAM, EU H2020
FET Proactive project RySQ and Scalable Ion-Trap Quantum Network (SciNet). 
Work at Harvard is supported through NSF, CUA, AFOSR Muri and the Vannevar Bush Faculty Fellowship.
AWG acknowledges funding from the National Research Foundation and
the Ministry of Education of Singapore. SC acknowledges support from Kwanjeong Educational Foundation.
HP is supported by the NSF through a grant for the ITAMP at Harvard University and the Smithsonian Astrophysical Observatory.

\appendix

\section{Lattice Corrections}\label{app:correction}

\begin{figure}[b]
\centering \includegraphics[width=.8\columnwidth]{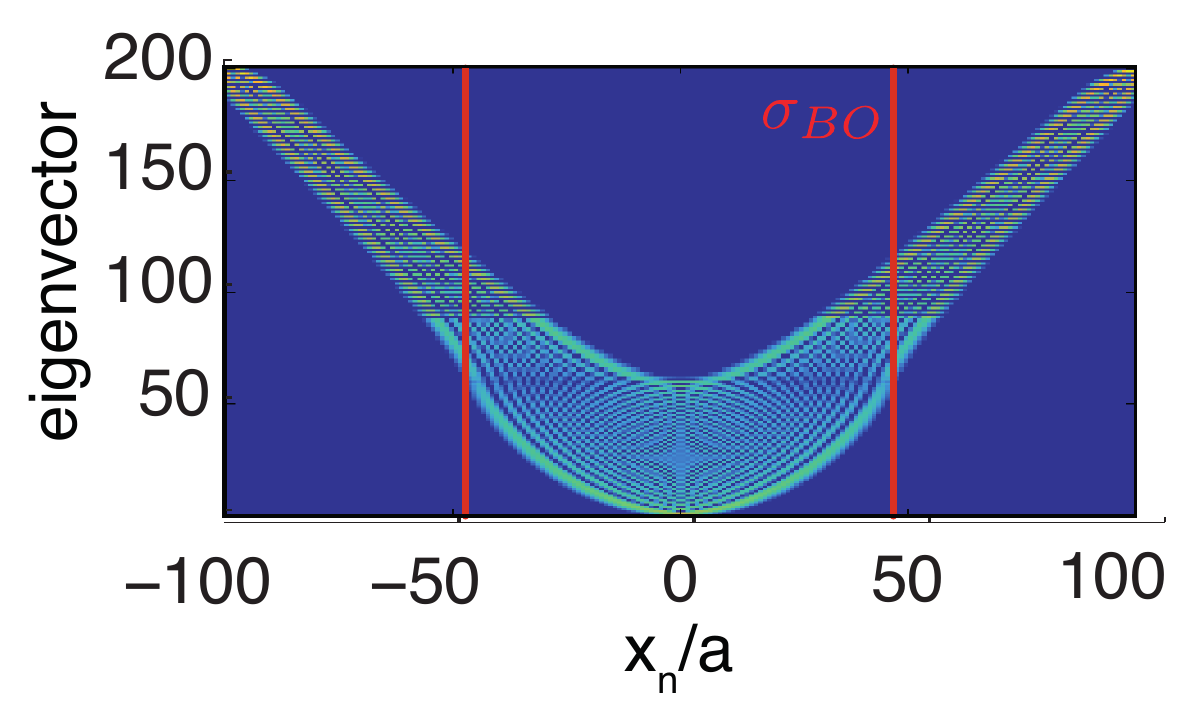}
\caption{{\small  Bloch oscillations: eigenfunctions of Hamiltonian \protect\eqref{eq:Hlattice}
illustrating Wannier-Stark localized eigenfunctions at the edge of
the lens (see text).}}
\label{fig:bloch} 
\end{figure}

In this Appendix we discuss dephasing and Bloch oscillations on the lattice and derive Eq.~\eqref{eq:optimum} for the optimal lens strength and Eq.~\eqref{eq:scaling} for the scaling of the final width.

{\it Dephasing:}
The optimal potential strengths $v_{c}$ and optimal pulse strength $\phi_{c}$ of Eq.~\eqref{eq:optimum} for the `thick' and `thin' lens, respectively, can be derived from the Bloch-band dispersion relation \mbox{$\epsilon(k)=2J[1-\cos(ka)]$} and its 
deviations from the quadratic dispersion relation \mbox{$\epsilon(k)-J(ka)^{2}\approx J(ka)^{4}/12+\mathcal{O}(ka)^6$}.
If these deviations become of the order of the inverse focusing
time, i.e. $J(ka)^{4}/12\sim t_{{\rm foc}}^{-1}(v_{0})$, then plane wave eigenstates will dephase during the focusing dynamics.
This happens for parts of the Wigner function exceeding a critical momentum
$k_{c}a=[2304v_{0}/(\pi^{2}J)]^{1/8}$ for the `thick' lens and $k_{c}a=(24\phi_{0})^{1/4}$ for the `thin' lens setup.
During focusing the distribution of momentum states populated will become broader with the largest width in momentum space, i.e. $k_{f}\sim1/\sigma_{f}$, at the focusing time. This limits the minimum final width and restricts the 
the `lens potential' to values below
\mbox{$v_{c}$} for the `thick' lens, as well as the critical pulse strength $\phi_{c}$
for the `thin' lens. 

{\it Bloch oscillations:}
The  onset of Bloch oscillations at $\sigma_{\rm BO}$ of Eq.~\eqref{eq:sbo} and the corresponding critical potential
strength $v_{\rm BO}$ can be understood in a semi-classical model for a particle in a quadratic potential with a Bloch-band dispersion relation, following the (semi-classical) equations of motion $\dot{x}=2Ja\sin( k a)$ and  ${\dot{k}}=-2v_{0}{x}$ for position and momentum, respectively. These equations are equivalent to a motion of a classical particle with a quadratic dispersion relation in a modified potential
\mbox{$V_{\rm BO}(x)=[(2v_{0}^{2}x_{0}^{2}-4v_{0}J)a^{2}x^{2}-(v_{0}^{2}a^{2})x^{4}]/2$}.
Depending on its initial position $x_{0}$ being smaller or larger than $\sigma_{\rm BO}$ the particle will either experience a single well or a double well potential. Fig.~\ref{fig:dynamics}(d) shows the corresponding eigenfunctions
of the lattice Hamiltonian~\eqref{eq:Hlattice}. Eigenfunctions which
have an extension less than $\sigma_{\rm BO}$ are well described
by discretized harmonic oscillator eigenfunctions centered around
the origin, while eigenfunctions with an extension larger than $\sigma_{\rm BO}$
start to get localized at the minima of the double wells of $V_{\rm BO}(x)$.
Thus, Bloch oscillations start to dominate the focusing dynamics at
a critical potential strength, $v_{{\rm BO}}=4J(a/\sigma_{0})^{2}$,
which is indicated as the red dashed line in Figs.~\ref{fig:corr}(a).

\section{Rydberg interaction between $^{87}$Rb atoms in $n_{1}S_{1/2}$ and
$n_{2}S_{1/2}$ Rydberg states}

\label{app:vdw}

\begin{figure*}[bt]
\centering \includegraphics[width=0.32\textwidth]{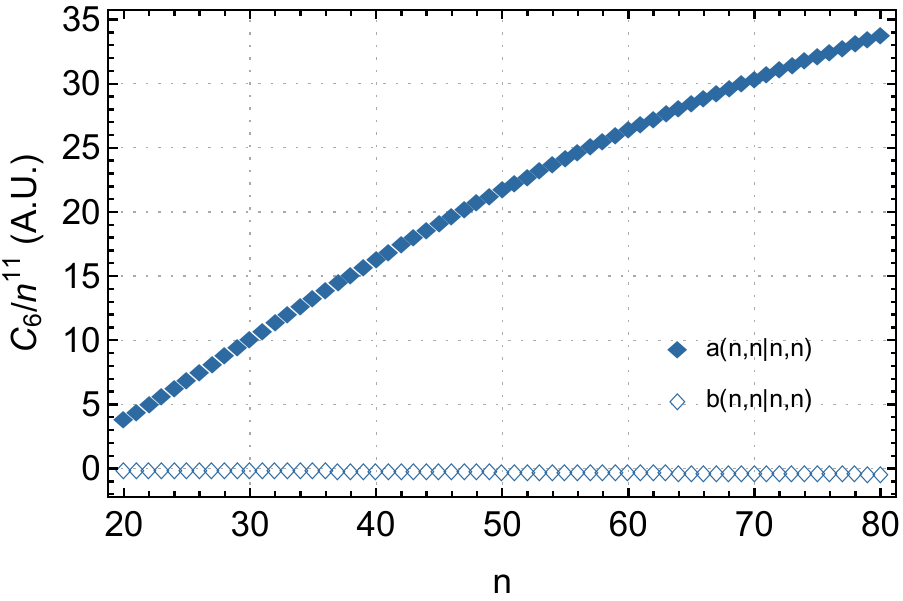} 
\includegraphics[width=0.32\textwidth]{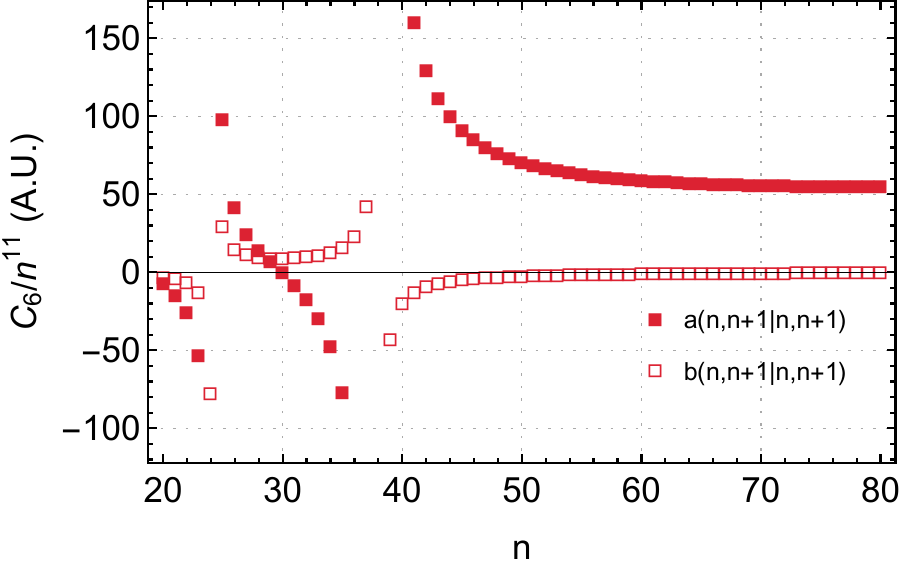}
\includegraphics[width=0.32\textwidth]{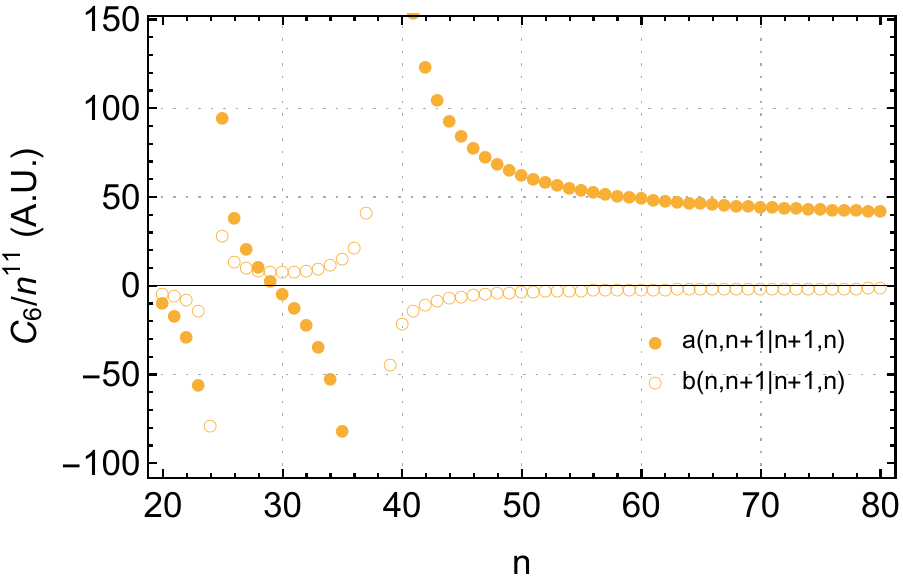} 
\caption{{\small{}{}{Generalized vdW coefficients of Eq.~\eqref{eq:Hab}
as a function of principal quantum number $n$ for the $nS_{1/2}$
and $(n+1)S_{1/2}$ Rydberg states.}}}
\label{fig:abcoeff} 
\end{figure*}

For distances large enough, such that the dipole interaction matrix
element between two $S$-states and two $P$-states is larger than
the energy difference $\Delta_{F}$ between these pair states, i.e.,
$V_{{\rm dip}}>\Delta_{F}$, we can treat vdW interactions perturbatively.
The vdW interaction Hamiltonian between $n_{1}S_{1/2}$ and $n_{2}S_{1/2}$
Rydberg states can be described by a 16$\times$16 matrix of the form 
\begin{widetext}
\begin{equation}
H_{{\rm vdW}}=\left(\begin{array}{cccc}
M(n,n|n,n) & 0 & 0 & 0\\
0 & M(n,n'|n,n') & M(n,n'|n',n) & 0\\
0 & M(n',n|n',n) & M(n',n|n',n) & 0\\
0 & 0 & 0 & M(n',n'|n',n')
\end{array}\right)\label{eq:C6mat}
\end{equation}
The vdW coefficients are given by 
\begin{equation}
\begin{split}M(n_{1},n_{2}|n_{3},n_{4})=\langle n_{1}S_{1/2}m_{1},n_{2}S_{1/2}m_{2}|H_{{\rm vdW}}|n_{3}S_{1/2}m_{3},n_{4}S_{1/2}m_{4}\rangle\end{split}
\end{equation}
is a 4$\times$4 matrix in the subspace of Zeeman levels $m=\pm1/2$
and the vdW interaction operator 
\begin{equation}
H_{{\rm vdW}}=\sum_{n_{\alpha},j_{\alpha},m_{\alpha}}\sum_{n_{\beta},j_{\beta},m_{\beta}}\frac{V_{{\rm dd}}|n_{\alpha}P_{j_{\alpha}}m_{\alpha},n_{\beta}P_{j_{\beta}}m_{\beta}\rangle\langle n_{\alpha}P_{j_{\alpha}}m_{\alpha},n_{\beta}P_{j_{\beta}}m_{\beta}|V_{{\rm dd}}}{E_{n_{1}}+E_{n_{2}}-E_{n_{\alpha}}-E_{n_{\beta}}},
\end{equation}
coupling $S$-states with energy $E_{n_{1}}$ and $E_{n_{2}}$ to
intermediate $P$-states with energies $E_{n_{\alpha}}$ and $E_{n_{\beta}}$
via dipole-dipole interactions 
\[
V_{{\rm dd}}(\mathbf{r})=-{\sqrt{\frac{24\pi}{5}}}\frac{1}{r^{3}}\sum_{\mu,\nu}C_{\mu,\nu;\mu+\nu}^{1,1;2}Y_{2}^{\mu+\nu}(\vartheta,\varphi)^{*}d_{\mu}^ {}d_{\nu}^ {}.
\]
Here, $\mathbf{d}$ is the atomic dipole operator and $\mathbf{r}=(r,\vartheta,\varphi)$
is the vector between the two atoms in spherical coordinates. With
$d_{\mu}$ we denote the $\mu$-th spherical components ($\mu,\nu\in\{-1,0,1\}$)
of the atomic dipole operator, $C_{m_{1},m_{2};M}^{j_{1},j_{2};J}$
are Clebsch-Gordan coefficients and $Y_{l}^{m}$ are spherical harmonics.
Using Wigner-Eckart's theorem the vdW interactions can be split up
in an angular and radial part 
\begin{equation}
M(n_{1},n_{2}|n_{3},n_{4})=\frac{1}{r^{6}}\left[a(n_{1},n_{2}|n_{3},n_{4})\mathbb{1}_{4}+b(n_{1},n_{2}|n_{3},n_{4})\mathcal{D}_{0}\right],\label{eq:Hab}
\end{equation}
with generalized isotropic and anisotropic vdW coefficients 
\begin{equation}
\begin{split}a & =\frac{1}{81}\left[7C_{6}^{(1)}+25C_{6}^{(2)}+11\left(C_{6}^{(3)}+C_{6}^{(4)}\right)\right],\\
b & =\frac{1}{27}\left[C_{6}^{(3)}+C_{6}^{(4)}-C_{6}^{(1)}-C_{6}^{(2)}\right],
\end{split}
\end{equation}
and $\mathbb{1}_{4}$ the 4$\times$4 identity matrix and

\begin{equation}
\mathcal{D}_{0}=\left(\begin{array}{cccc}
\cos(2\vartheta) & e^{-i\varphi}\sin(2\vartheta) & e^{-i\varphi}\sin(2\vartheta) & 2e^{-2i\varphi}\sin^{2}(\vartheta)\\
e^{i\varphi}\sin(2\vartheta) & \frac{2}{3}-\cos(2\vartheta) & -\cos(2\vartheta)-\frac{5}{3} & -e^{-i\varphi}\sin(2\vartheta)\\
e^{i\varphi}\sin(2\vartheta) & -\cos(2\vartheta)-\frac{5}{3} & \frac{2}{3}-\cos(2\vartheta) & -e^{-i\varphi}\sin(2\vartheta)\\
2e^{2i\varphi}\sin^{2}(\vartheta) & -e^{i\varphi}\sin(2\vartheta) & -e^{i\varphi}\sin(2\vartheta) & \cos(2\vartheta)
\end{array}\right)
\end{equation}
\end{widetext}

written in the basis $\{|\setbox0=\hbox{-}\vcenter{\hrule width\wd0height\the\fontdimen8\textfont3}\tfrac{1}{2}\setbox0=\hbox{-}\vcenter{\hrule width\wd0height\the\fontdimen8\textfont3}\tfrac{1}{2}\rangle,|\setbox0=\hbox{-}\vcenter{\hrule width\wd0height\the\fontdimen8\textfont3}\tfrac{1}{2}\tfrac{1}{2}\rangle,|\tfrac{1}{2}\setbox0=\hbox{-}\vcenter{\hrule width\wd0height\the\fontdimen8\textfont3}\tfrac{1}{2}\rangle,|\tfrac{1}{2}\tfrac{1}{2}\rangle\}$
of Zeeman states in the $j={1/2}$ Rydberg manifold and accounting
for the anisotropy and mixing between the Zeeman sublevels. With $C_{6}^{(\nu)}$
we denote the radial part of the matrix elements 
\begin{equation}
C_{6}^{(\nu)}(n_{1},n_{2}|n_{3},n_{4})=\sum_{n_{\alpha},n_{\beta}}\frac{\mathcal{R}_{1}^{\alpha}\mathcal{R}_{2}^{\beta}\mathcal{R}_{3}^{\alpha}\mathcal{R}_{4}^{\beta}}{\delta_{\alpha\beta}}
\end{equation}
which accounts for the overall strength of the interaction and is
independent of the magnetic quantum numbers. Here, $\mathcal{R}_{i}^{k}=\int drr^{2}\psi_{n_{i},\ell_{i},j_{i}}(r)^{*}r\,\psi_{n_{k},\ell_{k},j_{k}}(r)$
is the radial integral and $\nu$ accounts for the four channels to
intermediate $P_{j}$ states.

Figure~\ref{fig:abcoeff} shows the numerically calculated $a$ and
$b$ coefficients corresponding to the different blocks in Eq.~\eqref{eq:C6mat}
as a function of the principal quantum number $n$. Both $a$ and
$b$ show two F\"orster resonances around $n\approx24$ and $n\approx38$
where the channels to $\{nP_{3/2},nP_{1/2}\}$ and $\{nP_{3/2},nP_{3/2}\}$
states become close in energy, respectively. Apart from these resonances
the anisotropy coefficient $b$ is several orders smaller then the
diagonal coefficient $a$ which allows to safely neglect mixing of
Zeeman sublevels and results in an almost perfect isotropic interaction.

For two atoms initially in the $|n_{1}S_{1/2},1/2\rangle$ and $|n_{2}S_{1/2},1/2\rangle$
states this allows to reduce the dynamics to the four states $S_{1/2}$
states with principal quantum numbers $|n_{1},n_{1}\rangle$, $|n_{1},n_{2}\rangle$,
$|n_{2},n_{1}\rangle$ and $|n_{2},n_{2}\rangle$ and magnetic quantum
number $1/2$. The corresponding Hamiltonian restricted to this basis
has the form 
\begin{widetext}
\begin{equation}
H_{{\rm vdW}}=\frac{1}{r^{6}}\left(\begin{array}{cccc}
a(n_{1},n_{1}|n_{1},n_{1}) & 0 & 0 & 0\\
0 & a(n_{1},n_{2}|n_{1},n_{2}) & a(n_{1},n_{2}|n_{2},n_{1}) & 0\\
0 & a(n_{2},n_{1}|n_{2},n_{1}) & a(n_{2},n_{1}|n_{2},n_{1}) & 0\\
0 & 0 & 0 & a(n',n'|n',n')
\end{array}\right)
\end{equation}
\end{widetext}

with $a$ denoted as $C_{6}$ in the main text and plotted in Fig.~\ref{fig:Rydberg}(a)
as a function of the principal quantum number $n$.

For the long-range Hamiltonian of Eq.~\eqref{eq:RydSpin} (cf. Fig.~\ref{fig:Rydberg}(c))
next-nearest neighbor hopping is around 10\% of the nearest neighbor
hopping element. In Fig.~\ref{fig:nn_ato_comp}  we compare the performance of a spin lens implemented with the potentials arising from the Rydberg interactions, compared to the case with ideal nearest neighbor hopping. Our numerical results indicate that long-range
hopping terms slightly increase the speed of the scheme and decrease the final width.

\end{document}